\documentclass{mn2e}
\usepackage{epsf}
\usepackage{times} 

\def\j0658{J06587$-$5558}

\title{J06587$-$5558 --- A Very Unusual Polarised Radio Source}
 
\author[H.\ Liang et al.]
{H.\ Liang$^1$, R.\ D.\ Ekers$^2$, R.\ W.\ Hunstead$^3$, E.\ E.\ Falco$^4$, P.\ Shaver$^5$\\
$^1$ Physics Dept., University of Bristol, Tyndall Avenue, 
Bristol BS8 1TL, England\\
$^2$ Australia Telescope National Facility, CSIRO, PO Box 76, Epping
NSW 1710, Australia\\
$^3$ School of Physics, University of Sydney, NSW 2006, Australia\\
$^4$ Smithsonian Institution, Whipple Observatory, 670 Mt. Hopkins
Road, P.O. Box 97, Amado, AZ 85645 USA \\
$^5$ European Southern Observatory, Karl Schwarzchild-Strasse 2, 85748
Garching bei M\"unchen, Germany\\
}
\date{}
 
\pagerange{\pageref{firstpage}--\pageref{lastpage}}

\pubyear{}
 
\begin{document}
 
\maketitle
 
\label{firstpage}
 
\begin{abstract}
We have found a peculiar radio source \j0658 in the field of one of
the hottest known clusters of galaxies 1E0657$-$56. It is slightly
extended, highly polarised (54\% at 8.8\,GHz) and has a very steep
spectrum, with $\alpha \sim -1$ at 1.3\,GHz, steepening to $\sim -1.5$
at 8.8\,GHz ($S\propto \nu^{\alpha}$).  No extragalactic sources are
known with such high integrated polarisation, and sources with spectra
as steep as this are rare. 
In this paper, we report the unusual properties of the source \j0658
and speculate on its origin and optical identification.

\end{abstract}
 

\section{INTRODUCTION}
 
\begin{figure*}
\epsfxsize 400pt \epsfbox{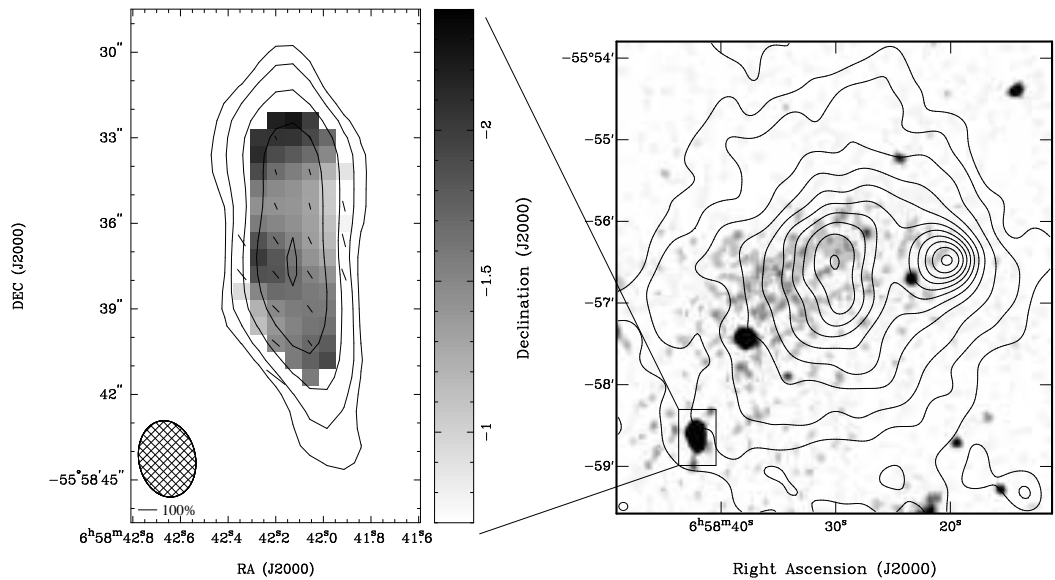}
\caption[]{{\it Left:} 4.8\,GHz total intensity contour map overlaid
on a grey scale spectral index (4.8--8.8\,GHz) image of \j0658.  The
contours are $(3,6,12,24,48)\times \sigma$, where $\sigma=40 \mu$Jy is
the rms noise in the image. The vectors represent the observed
E-vectors of the polarised emission at 4.8\,GHz and have not been
rotated back to their intrinsic position angles. The length of each
vector corresponds to the fractional linear polarisation; the length
of the vector corresponding to 100\% polarisation is given in the
bottom left hand corner together with the synthesised beam
($2.7''\times 2.0''$). The possible optical counterpart discussed in
section 6 is marked with a cross.  {\it Right:} A radio/X-ray overview
of the field of the cluster 1E~0657$-$56. The grey scale radio image
at 1.3\,GHz has a resolution of $6.5''\times 5.9''$; the beam is shown
in the bottom left hand corner. The X-ray contours are from a {\it
ROSAT\/} HRI image smoothed with a Gaussian of $10''$ FWHM. The
contour levels are 0.35, 0.45, 0.6, 0.8, 1.0, 1.2, 1.4, 1.6, 1.8, 2.0,
2.2\,HRI cts\,s$^{-1}$.  \label{hriradio} }
\end{figure*}

The cluster 1E~0657$-$56 at $z\sim 0.296$ is in many ways
exceptional. It has a high X-ray luminosity $L_{\rm bol}\sim 1.4\times
10^{46}$~erg~s$^{-1}$, and it is a contender for being the hottest
known cluster ($kT_{X}\sim 14.5$\,keV; Tucker et al.\ 1998, Liang et
al.\ 2000). The ROSAT PSPC/HRI images show that the cluster has two
distinct X-ray subclumps indicating on-going merging activity
(Fig.~\ref{hriradio}). Owing to its high X-ray luminosity, the cluster
was chosen for the detection of the Sunyaev-Zel'dovich effect with the
Swedish ESO Submillimeter Telescope (SEST; Andreani et al.\ 1999). The
SEST detection of the Sunyaev-Zel'dovich effect in the cluster
prompted subsequent attempts to confirm the detection at centimetre
wavelengths at the Australia Telescope Compact Array (ATCA). However,
any Sunyaev-Zel'dovich decrement at centimetre wavelengths was masked
by the presence of one of the most powerful cluster-wide diffuse radio
haloes known (Liang et al.\ 2000).  The radio halo is believed to be
of synchrotron emission and has an extent and morphology similar to
the X-ray emission.  When searching for possible polarisation in the
radio halo, we found an unusual highly polarised discrete source,
\j0658, about 2.5 arcmin from the cluster centre.

We discuss the nature of the source in the following sections. In
section 2, we briefly describe the radio properties of the source; in
section 3, we explore the possibilities of it being Galactic or
extragalactic; sections 4 and 5 discuss whether it is inside or behind
the cluster; section 6 gives the results of deep imaging with the VLT,
and the conclusions are given in section 7.

Throughout the paper we will assume $\Omega_{M}=0.3$,
$\Omega_{\Lambda}=0.7$ and $H_{0}=75$\,km\,s$^{-1}$\,Mpc$^{-1}$.
 
\section{A PECULIAR RADIO SOURCE}
 
\begin{figure}
\epsfxsize 200pt \epsfbox{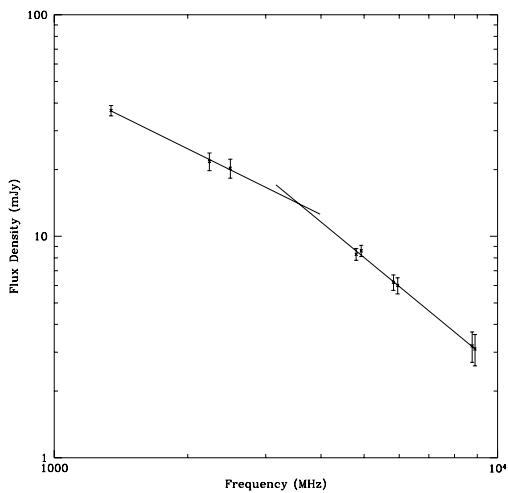}
\caption{Total radio flux density versus frequency. The low frequency
spectral index is $-1$ and the high frequency index is
$-1.5$, implying a spectral break in the region of 3--4 GHz.\label{sp}}
\end{figure}

Radio observations centred on the cluster 1E~0657$-$56 were obtained
at the ATCA using various antenna configurations and observing
frequencies.  The observations were aimed initially at the attempted
detection of the Sunyaev-Zel'dovich effect, subtraction of confusing
radio sources and subsequent studies of the radio halo
emission. Details of the radio observations are given in Liang et al.\
(2000).  The peculiar radio source \j0658 stands out in the
polarisation maps with $\sim 54$\% linear polarisation at 8.8\,GHz. It
is significantly depolarised with $<0.5$\% polarisation at
1.3\,GHz. No circular polarisation was detected, with an upper limit
of 0.5\% at 1.3\,GHz. The source has a steep spectrum, with a spectral
index (defined by $S\propto \nu^{\alpha}$) of $\alpha^{1.3\,\rm
GHz}_{2.2\,\rm GHz} = -1.0$, steepening at higher frequencies to reach
$\alpha^{4.8\,\rm GHz}_{8.8\,\rm GHz} = -1.5$ (Fig.~\ref{sp}). The
properties of the source are summarised in Table~\ref{tb1}.

The spectrum and polarisation suggest that the emission is
synchrotron. Synchrotron emission is intrinsically highly polarised
(up to $\sim 70$\%), but observed polarisations are usually weak due
to various depolarisation effects. While a small region (e.g. part of
a jet) of an extragalactic source can be highly polarised, so far no
extragalactic radio source is known with such a high {\it
integrated\/} linear polarisation (e.g.\ de Zotti et al.\ 1999), and
only a few percent of the sources found at high radio frequencies have
spectra as steep as \j0658 (e.g.\ Richards 2000).

\section{GALACTIC OR EXTRAGALACTIC?}
Radio emission with such high polarisation and such a steep spectrum is
normally seen only from pulsars; in fact, the millisecond pulsars were
discovered this way.  A high resolution follow-up observation obtained
with the ATCA in 1997 November using the 6C configuration at 4.8 and
8.8\,GHz showed that the source was elongated by $\sim 15''\times
4.5''$ (Figs~\ref{hriradio}~\&~\ref{3cm}). The deconvolved
elliptical Gaussian size of the source is $\sim 7.8''\times
1.8''$ FWHM at all frequencies from 1.3 to 8.8\,GHz. Since the source is
extended, and by a similar amount at all frequencies, it cannot be a
pulsar, or even a scattered pulsar. The agreement in angular size also
indicates that there is no significant systematic loss of
short-spacing flux at high frequencies, so the steepening of the
spectrum in Fig.~\ref{sp} is real.

If Faraday rotation is caused by the external media between the source
and the observer, then the position angle of the electric vector
($\Phi$) is related to wavelength ($\lambda$) by
\begin{equation}
\Phi=\Phi_{0}+{\rm (RM)} \lambda^{2}
\label{eq1}
\end{equation}
where RM is the rotation measure.  The position angle versus
$\lambda^{2}$ relation is well fitted by a straight line 
to five frequencies spanning the range 2.2--8.9 GHz.  Between 4.8 and
8.8\,GHz, at a resolution of $2.7''\times 2.0''$, we obtain an average
rotation measure of $-266\pm 37$\,rad\,m$^{-2}$ across the source.
Figure~\ref{3cm} shows the rotation measure across the source in
greyscale. For comparison, the Galactic rotation measure in the
direction of the source is $\sim +64$\,rad\,m$^{-2}$ (Simard-Normandin
et al. 1981), or $+36\pm 28$\,rad\,m$^{-2}$ (J.\ Han, private
communication). Hence, the rotation measure in \j0658 is unlikely to
be Galactic in origin, and this, together with the absence of a clear
optical counterpart (see section 6), argues that the source must be
extragalactic.

\section{INSIDE THE CLUSTER?}

\begin{figure}
\epsfxsize 200pt \epsfbox{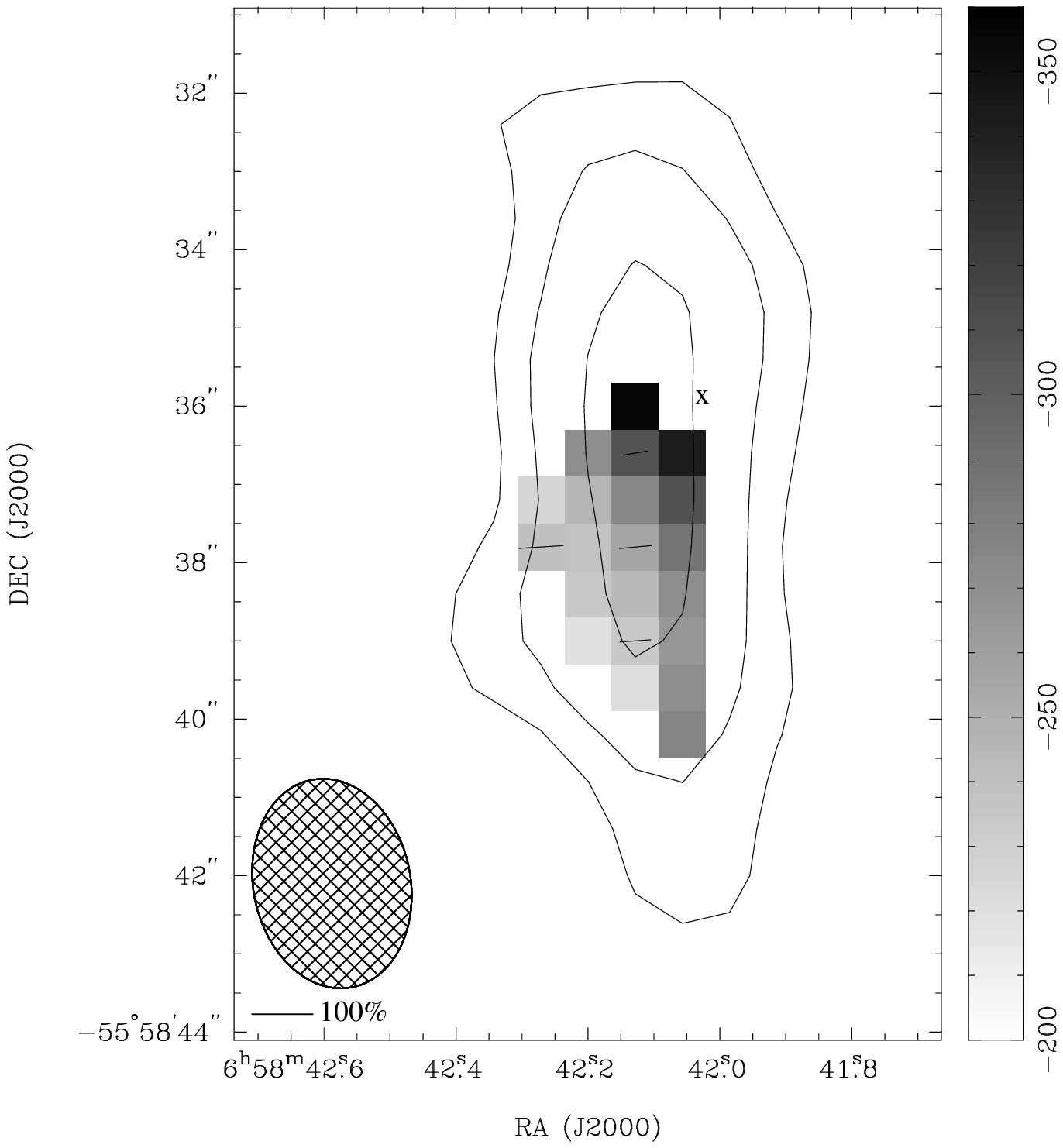}
\caption{A total intensity contour image at 8768\,MHz overlaid on a
greyscale rotation measure image. The E-vectors are shown rotated back
to the intrinsic position angles of the linearly polarised emission at
8768\,MHz, and their lengths give the percentage linear polarisation
(the vector length corresponding to 100\% polarisation is shown in the
bottom left-hand corner). The contours are $(3,6,12)\times \sigma$,
where the rms noise of the image is $\sigma=50 \mu$Jy. The beam is
shown in the bottom left-hand corner ($2.7''\times 2.0''$) and the
location of the possible optical counterpart (Sec. 6) is marked with a
cross. \label{3cm}}
\end{figure}

\begin{table}
\caption{Properties of the peculiar radio source \label{tb1}}
\begin{tabular}{ll}
\hline
\hline
Position (J2000) & RA = 06 58 42.1 \\
                 & Dec = $-$55 58 37 \\
Total extent     & $15''\times 4.5''$ \\ 
Linear size      & $62\times 19$\,kpc ($z=0.296$) \\
                 & $103  \times 31$\,kpc ($z=3.5$) \\
$S_{1.3}$ (note $a$)       & $37\pm 2$\,mJy \\[1mm]
$L_{1.4}$ (note $b$) & $9\times 10^{24}$\,W\,Hz$^{-1}$ ($z=0.296$) \\
                & $3.5\times 10^{27}$\,W\,Hz$^{-1}$ ($z=3.5$) \\
$\alpha_{4.8}^{8.8}$ (note $c$) & $-1.5\pm 0.3$ \\[1mm]
$\alpha_{1.3}^{2.2}$ (note $d$) & $-1.0\pm 0.1$ \\
Linear pol.(8.8\,GHz) & $(54\pm 11)$\% \\
Linear pol.(4.8\,GHz) & $(45\pm  9)$\% \\
Linear pol.(2.2\,GHz) & $(4.3\pm 0.5)$\% \\
Linear pol.(1.3\,GHz) & $<0.5$\% \\
RM (note $e$)    & $-266\pm 37$ radians~m$^{-2}$ \\
PA$_{0}$ (note $f$) & $-84^{\circ}\pm 2^{\circ}$\\ 
\hline
\end{tabular}

\medskip
$^{a}$ Total flux density at 1.3\,GHz\\
$^{b}$ Total radio luminosity at 1.4\,GHz (rest frame) \\
$^{c}$ Average spectral index across the source between 4.8 and 8.8\,GHz\\
$^{d}$ Average spectral index across the source between 1.3 and 2.2\,GHz\\
$^{e}$ Rotation measure across the source with a resolution of $2.7''\times 2.2''$\\
$^{f}$ Intrinsic position angle of the E-vector\\
\end{table}

\begin{figure*}
\centerline{\epsfxsize 0.5\textwidth \epsfbox{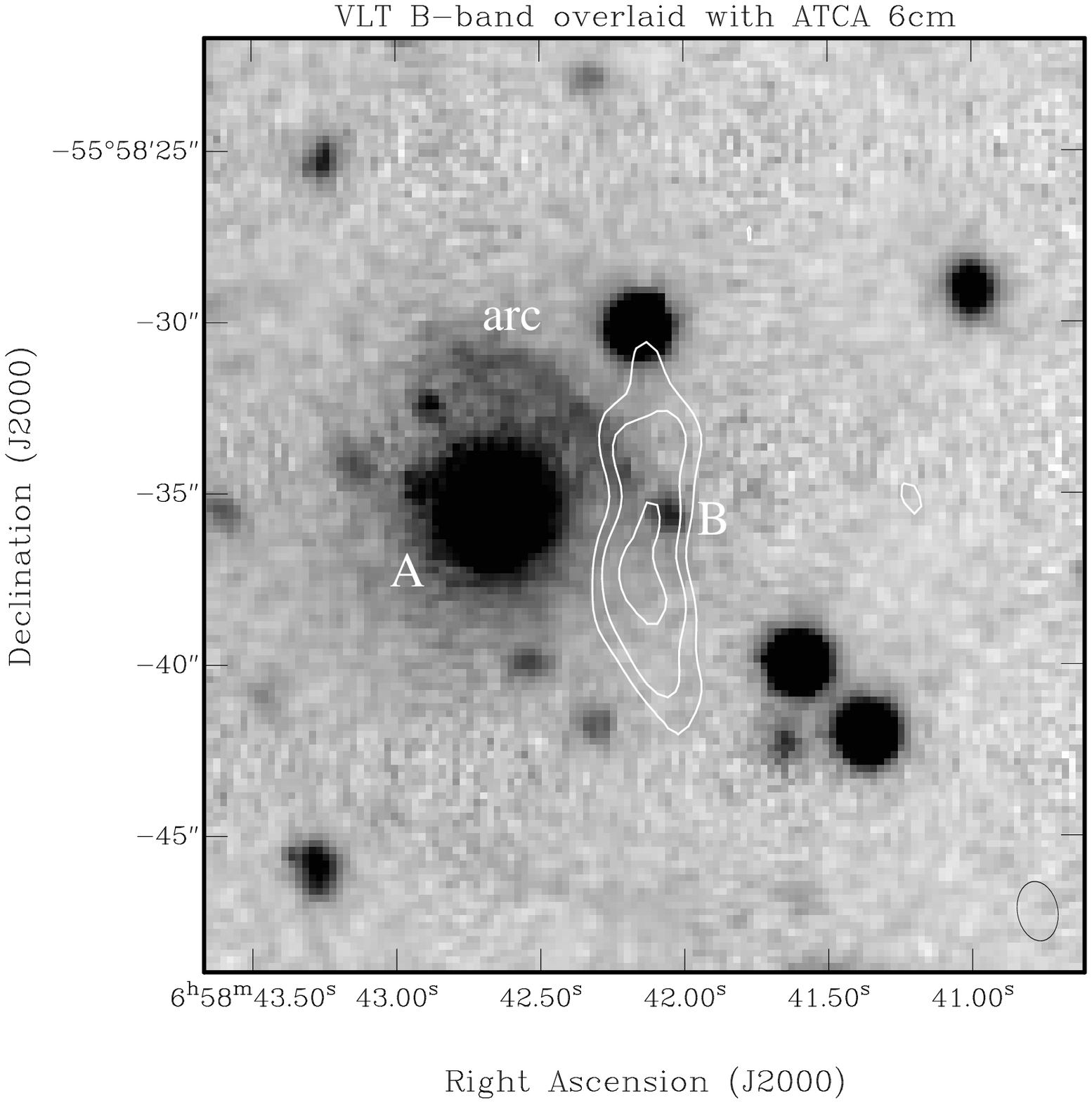} \epsfxsize 0.5\textwidth \epsfbox{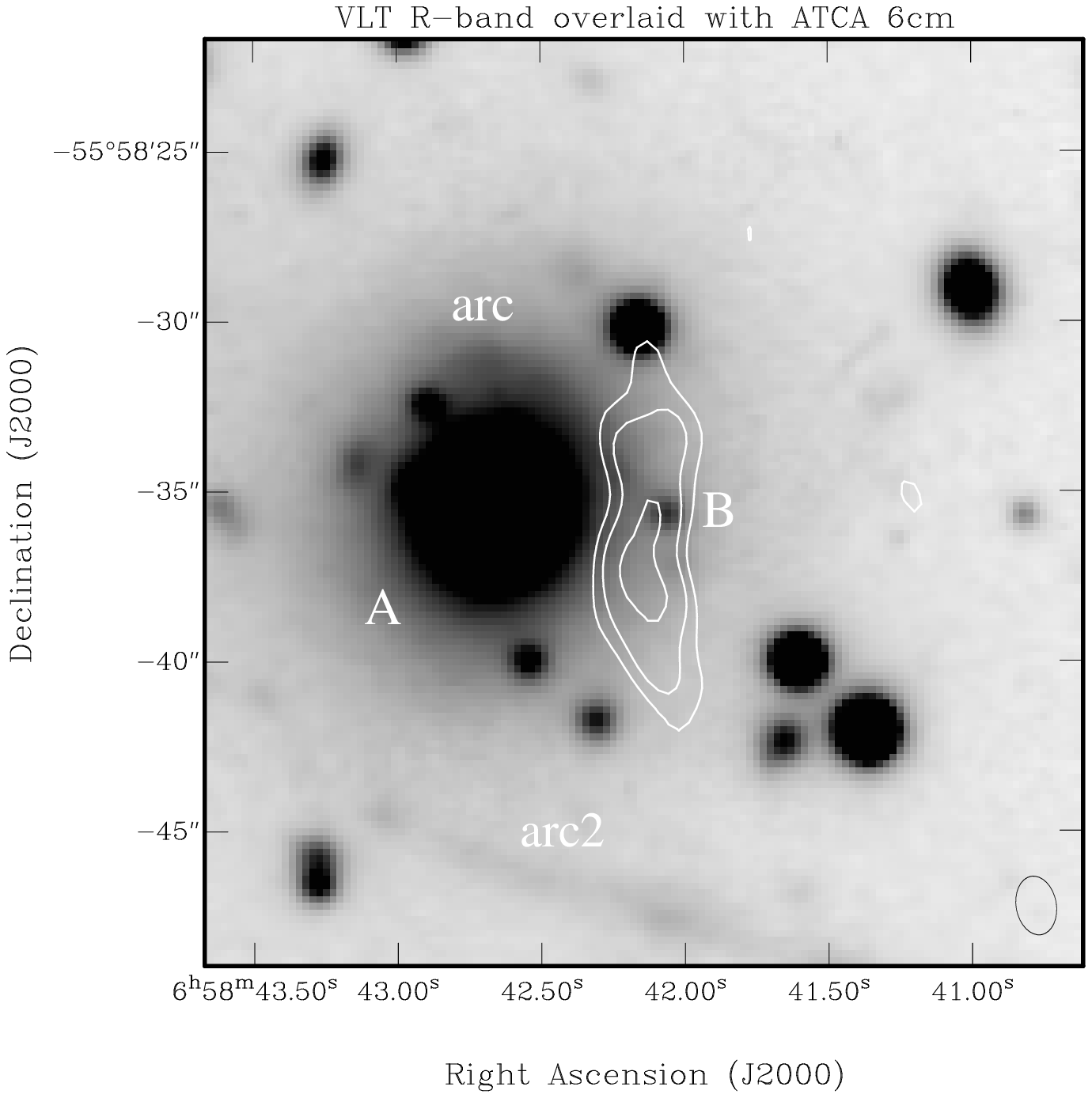}}
\caption{A high-resolution 4800\,MHz contour map overlaid on VLT
B-band ({\it left}) and R-band ({\it right}) images; the measured
seeing was $0.75''$ FWHM. Galaxy A is a cluster member, while faint
galaxy B is the brightest object which falls within the radio source
envelope.  Two possible gravitational lens arcs associated with galaxy A
are marked.
Contours are at 0.2, 0.4, 0.8 mJy/beam, and the rms noise of the
image is $\sigma=65 \mu$Jy. The radio beam ($1.7''\times 1.2''$)
is shown in the lower right-hand corner of each image.
\label{vlt}}
\end{figure*}

The ESO 3.6-m telescope at La Silla was used to obtain 20-minute
exposures in both B- and I-band, as well as spectroscopy for the
surrounding objects in Oct. 1997 by P. Shaver and I. Hook. There is no
obvious optical identification for \j0658, but the bright elliptical
galaxy $5''$ to the east (galaxy A in Fig.~\ref{vlt}) was confirmed as
a cluster member ($z\sim 0.293$).

If the source is in the cluster then the lack of any optical
identification with cluster galaxies suggests that it may possibly be
a radio relic $62\times 19$\,kpc in extent.  Diffuse radio relics of
the type seen in A3667 and A2256 are known to be polarised with very
steep spectra, though without significant depolarisation.  For
instance, the relics (G and H) in A2256 are polarised at a $20$\%
level at 1.4\,GHz (R\"ottgering et al.\ 1994). However, these relics
are much larger in spatial extent ($\sim 1$\,Mpc) and more diffuse
than \j0658.

The source appears to be embedded in the outer parts of the radio
halo, with its long axis approximately tangential to the X-ray
contours. It is conceivable that \j0658 is just an enhanced part of
the radio halo where the local magnetic field is strong and well
organised, as in the case of a shock seen in the plane of
compression. Figure~\ref{3cm} shows that the magnetic field vectors
are well aligned with the direction of the source elongation over at
least the southern half of the source. The surface brightness of
\j0658 is roughly 100 times that of the halo, so the magnetic field in
the source would need to be at least 10 times greater than the large
scale cluster field, which is still reasonable.

\section{BEHIND THE CLUSTER?}

We have also searched in the near infrared in case the optical
emission is obscured by dust, or the object is at a redshift 
such that most of the optical emission above the 4000 \AA\ break has
been shifted into the infrared. We observed the field for 4.5 hrs on
1999 November 23 with the infrared camera CASPIR on the ANU 2.3m
telescope at Siding Spring Observatory. No infrared counterpart was
found to a limiting magnitude of H $\sim 20$.

If the source is behind the cluster, then it may be a radio galaxy at
high redshift. The alignment of the magnetic field with the elongation
of the source is consistent with that of large-scale jets. The
depolarisation and slight flattening of the spectrum near the position
of the possible optical counterpart (Fig.~\ref{hriradio}) may also be
consistent with a core-jet structure in which the core is often
depolarised with a flatter spectrum.  High-redshift radio galaxies are
known to have steep spectral indices ($\alpha<-1$). However, in a
polarisation study of a sample of high redshift radio galaxies,
Carilli et al.\ (1997) found a maximum integrated polarisation of
$\sim 20$\% at 8.2\,GHz. Hence, if this source is a high redshift
radio galaxy, it is exceptional in terms of its polarisation
properties.

Alternatively, \j0658 could also be the gravitationally lensed part of
a background jet since small parts of a jet can sometimes be highly
polarised and the more polarised parts tend to have a steeper
spectrum. If it is just part of the jet, then we do not expect to find
an optical identification at the position of the radio emission. The
cluster 1E0657$-$56 is one of the most massive clusters known; it can
act as an effective gravitational lens. To produce the orientation of
the source, it is necessary to have another lens to the other side of
the source. The elliptical cluster galaxy (A) to the east of the
source could act as such a counterweight.

\section{OPTICAL IDENTIFICATION FROM THE VLT}

VLT images of the field of \j0658 were obtained by Mehlert et al.\
(private communication) with UT1 in the B, g, R and I bands during the
FORS1 commissioning in December 1998. Exposure times were 10 minutes
and the measured seeing was $0.75''$ FWHM.  These images revealed a
faint possible identification (B), embedded in the halo of the galaxy
(A) to the east of the source (Fig.~\ref{vlt}), at 06 58 42.04, $-$55
58 35.8 (J2000).

The astrometric transfer to the deep VLT images was done using the
publicly available SuperCOSMOS scans of the red (IIIaF) sky survey
plate of Field 162 from the UK/AAO Schmidt telescope.  The formal fit
error on this plate is given as $0.13''$ rms in each coordinate.  The
accuracy of radio-optical registration was confirmed to be $\leq
0.2''$ by overlaying the full resolution ATCA 3 cm images of two
tailed cluster source in 1E 0657$-$56 on the VLT image.  The
displacement of galaxy B from the axis and centroid of the radio
emission ($\sim 0.6''$) is real, and therefore casts doubt on any possible
association.

To estimate fluxes in the VLT B, g, R and I bands, we extracted a
patch $\sim 26\arcsec\times26\arcsec$ ($128^2$ pixels) centred on the
nearby elliptical galaxy (A), and fitted a photometric model to the galaxy (A)
and to eight compact sources embedded in the extended galaxy profile, including
galaxy B. We selected an isolated, bright but unsaturated, star from
each image to define the point-spread function (PSF), and utilised the
{\sc imfit} package, written by B.\ McLeod, which has been used to
determine photometric models for lensed systems in the CASTLES project
(see, e.g., Leh\'ar et al. 2000). We used magnitude zero points kindly
provided by S.\ Seitz. The quality of the fits was high, yielding
estimated uncertainties of $\sim 0.05$ mag.  Apparent magnitudes of
galaxy B thus measured in B, g, R and I bands are listed in
Table~\ref{tb2} along with the upper limit in H band.

\begin{figure}
\epsfxsize 200pt \epsfbox{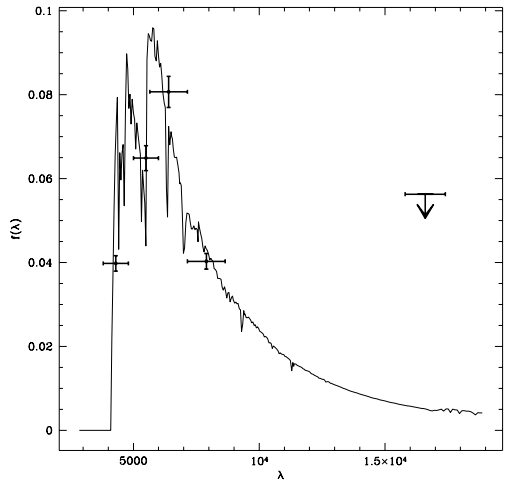}
\caption{A synthetic SED (solid curve) corresponding to an elliptical
galaxy at $z=3.5$ is plotted along with the observed broad-band fluxes
($f(\lambda)$ has units of erg\,s$^{-1}$\,cm$^{-2}$\,\AA$^{-1}$)
against the wavelength in the observer's frame of reference. The
vertical error bars correspond to the photometric errors and the
horizontal bars correspond to the width of the filters in wavelength.}
\label{sed}
\end{figure}

The steepness of the radio spectrum suggests that \j0658 is unlikely
to be a quasar or a Seyfert galaxy.  Radio galaxies are identified
with giant elliptical galaxies. It has been shown by Scarpa \& Urry
(2000) that there is no fundamental difference between the optical
properties of radio and non-radio emitting elliptical galaxies except
that the probability of an elliptical galaxy hosting a radio source is
proportional to the square of its optical luminosity.  Assuming that
galaxy B is an elliptical, we found a solution for the photometric
redshift, $z_{\rm ph}=3.5\pm 0.1$, using the public domain photometric
redshift program {\sc hyperz} (Bolzonela et al.\ 2000). A synthetic
Spectral Energy Distribution (SED) corresponding to an elliptical
galaxy (Bruzual \& Charlot 1993) was fitted to the broad band
magnitudes taking into account the dust absorption through the galaxy
itself using the reddening law of Calzetti et al. (2000), and the
continuum absorption by the Lyman forest using estimates for
Lyman-$\alpha$ and Lyman-$\beta$ line blanketing by Madau (1995) and
zero flux throughput below the Lyman limit. The SED fit to the observed
broad band flux is given in Fig.~\ref{sed}. At $z\sim 3.5$, the source
would have $M_{R} \sim -21.3$ and $L_{1.4\,\rm GHz} \sim 3.5\times
10^{27}$\,W\,Hz$^{-1}$, typical of a classical FR II radio galaxy.

Finally, we examine the significance of galaxy B being within the
radio contours of \j0658. In the small $\sim 26\arcsec\times26\arcsec$
patch centred on galaxy A, we found 10 faint objects in the B-band
image with magnitudes similar to galaxy B. Hence the probability of
finding an object like galaxy B within the second radio contour ($\sim
8\arcsec\times3\arcsec$) by chance is $\sim 36$\% which shows that the
coincidence is not of high significance.  

We also note that there are two arc-like features, one long straight
arc to the south and another arc to the north-west of galaxy A, a
confirmed cluster member. The long arc is seen in the R-band (`arc2'
in Fig.~\ref{vlt}) as well as the I-band image and it is most likely
to be a gravitational arc.  The other arc is most prominent in the
B-band image (`arc' in Fig.~\ref{vlt}). It could be a gravitational
arc or a tidal feature as a result of merging. If it is a
gravitational arc, then it makes the lensing scenario given at the end
of Sec. 5 more plausible since the radio contour would then be
traversing the caustics. If it is a merging signature, then it lends
support to the presence of shocks and hence the cluster relic scenario
(Sec. 4).

\begin{table}
\caption{Optical and near infrared magnitudes of galaxy B. The
magnitudes have been corrected for Galactic extinction. \label{tb2}}
\begin{tabular}{lllll}
\hline \hline \multicolumn{1}{c}{$m_{\rm B}$} &
\multicolumn{1}{c}{$m_{\rm g}$} & \multicolumn{1}{c}{$m_{\rm R}$} &
\multicolumn{1}{c}{$m_{\rm I}$} & \multicolumn{1}{c}{$m_{\rm H}$}\\
\hline\\ 24.75$\pm$0.05 & 23.84$\pm$0.05 & 22.79$\pm$0.05 &
22.92$\pm$0.05 & $>$20\\ \hline\\
\end{tabular}

%\medskip
%Note. -- The magnitudes have been corrected for Galactic extinction.
\end{table}

\section{CONCLUSIONS}
We have found an unusual extended radio source with a steep radio
spectrum and extreme integrated linear polarisation (Table~\ref{tb1}).
After eliminating the likelihood that the source is Galactic, we have
considered three possible explanations for these unusual properties:

\begin{itemize}
\item Gravitational lens\\ \j0658 is the lensed fragment of a jet
belonging to a distant radio galaxy.  This would explain the high
polarisation since small regions of radio galaxies are known to have
high polarisation. The proximity of the bright elliptical (A), the
presence of the rich cluster and the arcs in the immediate
surroundings lend support to this scenario.

\item $z=3.5$ radio galaxy with no lensing\\ The optical
identification (galaxy B) is offset from the radio central axis, and
the probability of a chance coincidence of galaxy B falling within the
radio contours of \j0658 is high. If galaxy B is not the true
identification then the true identification must be even fainter which
means that it is a radio galaxy at $z>3.5$. In either case, the high
polarisation is very unusual since high redshift radio galaxies are not
known to have such high polarisation (Carilli et al. 1997).

\item Cluster relic\\ The presence of the rich cluster, the nearby
bright elliptical galaxy (A) and the possible tidal feature shows that
\j0658 maybe in regions of shocks which favour the formation of
cluster relics. However, known cluster relics are larger in spatial
extent, more diffuse and without significant depolarisation.

\end{itemize}

\bigskip

We are grateful to Isobel Hook for taking the ESO 3.6-m data, Stella Seitz
for providing the VLT magnitude zero points, and Hans B\"ohringer for
obtaining a spectrum of one of the galaxies. We acknowledge the use of
the {\sc hyperz} photometric redshift program
(http://webast.ast.obs-mip.fr/hyperz/), and the use of the {\sc karma}
package (http://www.atnf.csiro.au/karma) for the overlays.

\end{document}